# Article information

## Article title
Hematoxylin and eosin stained oral squamous cell carcinoma histological images dataset

## Authors
Dalí F. D. dos Santos[*,a], Paulo R. de Faria[b], Adriano M. Loyola[c], Sérgio V. Cardoso[c], Bruno A. N. Travençolo[a], Marcelo Z. do Nascimento[a]

## Affiliations
[a] Faculty of Computer Science, Federal University of Uberlândia, Brazil
[b] Institute of Biomedical Science, Federal University of Uberlândia, Brazil
[c] Department of Oral and Maxillofacial Pathology, School of Dentistry, Federal University of Uberlândia, Brazil

## Corresponding author's email address and Twitter handle
dalifreire@gmail.com

## Abstract
Computer-aided diagnosis (CAD) can be used as an important tool to aid and enhance pathologists' diagnostic decision-making. Deep learning techniques, such as convolutional neural networks (CNN) and fully convolutional networks (FCN), have been successfully applied in medical and biological research. Unfortunately, histological image segmentation is often constrained by the availability of labeled training data once labeling histological images for segmentation purposes is a highly-skilled, complex, and time-consuming task. This paper presents the hematoxylin and eosin (H&E) stained oral cavity-derived cancer (OCDC) dataset, a labeled dataset containing H&E-stained histological images of oral squamous cell carcinoma (OSCC) cases. The tumor regions in our dataset are labeled manually by a specialist and validated by a pathologist. The OCDC dataset presents 1,020 histological images of size 640x640 pixels containing tumor regions fully annotated for segmentation purposes. All the histological images are digitized at 20x magnification.

## Specifications table

| | |
|---|---|
| **Subject** | Computer Science, Computer Vision, and Pattern Recognition. |
| **Specific subject area** | Tumor segmentation, H&E-stained histological images, Tumor histological images, and Oral biopsy images. |
| **Type of data** | Image. |
| **How the data were acquired** | The original images were captured as whole slide images (WSI) using the Slide Scanner Aperio AT2 at 20x magnification. A total of 1,020 image patches of size 640×640 pixels were randomly extracted from the captured WSIs. The tumor regions present in each image patch were hand-annotated for segmentation purposes by a specialist and fully validated by a pathologist. |
| **Data format** | Raw. |
| **Description of data collection** | The OCDC dataset was built using tissue specimens retrieved from OSCC-affected patients from the Department of Oral and Maxillofacial Pathology archives at the Federal University of Uberlândia. The OCDC dataset comprises 1,020 images of size 640×640 pixels containing normal and tumor areas. The tumor regions present in each image were hand-annotated for segmentation purposes and validated by a pathologist. The images are taken at 20x magnification. |
| **Data source location** | ● Institution: Department of Oral and Maxillofacial Pathology archives from the Federal University of Uberlândia<br>● City/Town/Region: Uberlândia, state of Minas Gerais<br>● Country: Brazil |
| **Data accessibility** | Repository name: Hematoxylin and eosin stained oral squamous cell carcinoma histological images dataset<br><br>Freire, Dalí; Faria, Paulo; Loyola, Adriano; Cardoso, Sergio; Travencolo, Bruno; do Nascimento, Marcelo (2022), "H&E-stained oral squamous cell carcinoma histological images dataset", Mendeley Data, v1<br><br>Data identification number: 10.17632/9bsc36jyrt.1<br><br>Direct URL to data: https://data.mendeley.com/datasets/9bsc36jyrt/1<br><br>Direct URL to the dataset in GitHub repository: https://github.com/dalifreire/tumor_regions_segmentation/tree/main/datasets/OCDC |
| **Related research article** | D.F.D. dos Santos, P.R. de Faria, B.A.N. Travençolo, M.Z. do Nascimento, Automated detection of tumor regions from oral histological whole slide images using fully convolutional neural networks. Biomedical Signal Processing and Control 69, 102921 (2021). https://doi.org/https://doi.org/10.1016/j.bspc.2021.102921, https://www.sciencedirect.com/science/article/pii/S1746809421005188 |

**Value of the data**

- An H&E-stained histological OSCC images dataset with pixel-level tumor annotations designed for segmentation purposes.
- Useful in the development of computational techniques for histological image segmentation to support pathologists in decision-making in cases of oral cavity-derived cancer.
- Can be used by computer science researchers to enhance and compare the segmentation results achieved by different segmentation methods.

## Objective

The availability of labeled training data often constrains the development of computational techniques for histological image segmentation once labeling histological images for segmentation purposes is highly skilled, complex, and time-consuming. OSCC is a kind of cancer that is scarcely studied because of the lack of available labeled training images. Deep learning models traditionally need large amounts of labeled data samples to be trained. The OCDC dataset allows to conduct of new studies on OSCC and can be used by other researchers for testing their computational approaches to OSCC region segmentation.

## Data description

The OCDC dataset consists of 1,020 H&E-stained histological images of size 640x640 pixels and their corresponding pixel-level tumor regions annotation masks. The 1,020 images and their corresponding labels were randomly split into 840 images to train and 180 images to test [3]. Fig. 2 illustrates the images and tumor regions annotation masks of the dataset. The OCDC dataset was built using tissue specimens collected from OSCC-affected patients. All OSCC cases were retrieved from the Department of Oral and Maxillofacial Pathology archives at the Federal University of Uberlândia between 2006 and 2013 under approval by the Committee on Research and Ethics of the Institution (CAAE number: 15188713.9.0000.5152).

The dataset images are split into 840 images to train and 180 images to test. The 1,020 images of the OCDC dataset contain both images of OSCC tumor regions and images of normal tissues like the serous salivary gland, connective tissue, mucous salivary gland, striated muscle, keratinized epithelial tissue, and oral mucosa. All the raw data are publicly available in a Mendeley data repository:

- https://data.mendeley.com/datasets/9bsc36jyrt/1

## Experimental design, materials and methods

To create the OCDC dataset, tissue samples obtained after surgical procedure in 15 OSCC-affected patients were digitized into 15 WSIs using the Slide Scanner Aperio AT2 (Leica Biosystems Imaging, Inc., Nussloch, Germany) coupled to a computer (Dell Precision T3600) at 20x magnification and pixel-level resolution of 0.5025µm x 0.5025µm. The digitized WSIs have different sizes – the larger one has almost three billion pixels (63,743 x 45,472 pixels) – and are stored in SVS [2] format using the RGB color space. H&E-stained histological WSIs are multi-gigapixel images created from tissue sections that contain many different types of cells, tissues, and structures, such as blood vessels, keratin, lymphocytes, glands, muscle, and tumor cells [1].

A total of 1,020 images of size 640 x 640 pixels were randomly extracted from these 15 WSIs and the tumor regions present in each image were hand-annotated by a specialist for segmentation purposes [3]. The tumor regions were annotated with the GNU Image Manipulation Program (GIMP) using a pen on a touch screen monitor. A pathologist validated the resulting pixel-level annotated images. This kind of task and a large amount of information in the images require very skilled professionals and a great effort to label the relevant structures to create this kind of image dataset. Fig. 1 illustrates the OSCC region labeling process from the OCDC dataset.

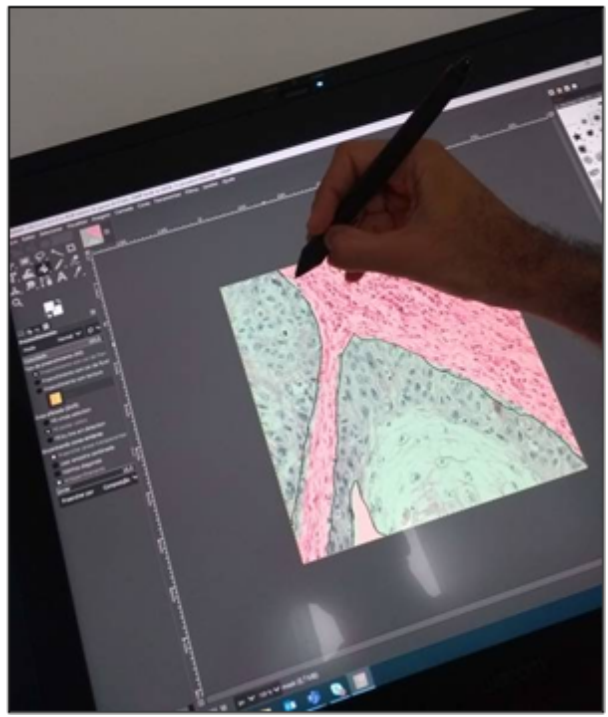

**Fig. 1.** The labeling process from the OCDC dataset: Hand-annotating tumor regions for image segmentation purposes (green and red areas indicate tumor regions and normal tissue, respectively).

The image dataset was split into two subsets: 840- and 180-image-containing training and test sets, respectively [3]. Representative examples from the 1,020 produced images are shown in Fig. 2. Fig. 2 (m-r) also shows the corresponding hand-annotated regions produced from the images presented in Fig. 2 (g-l).

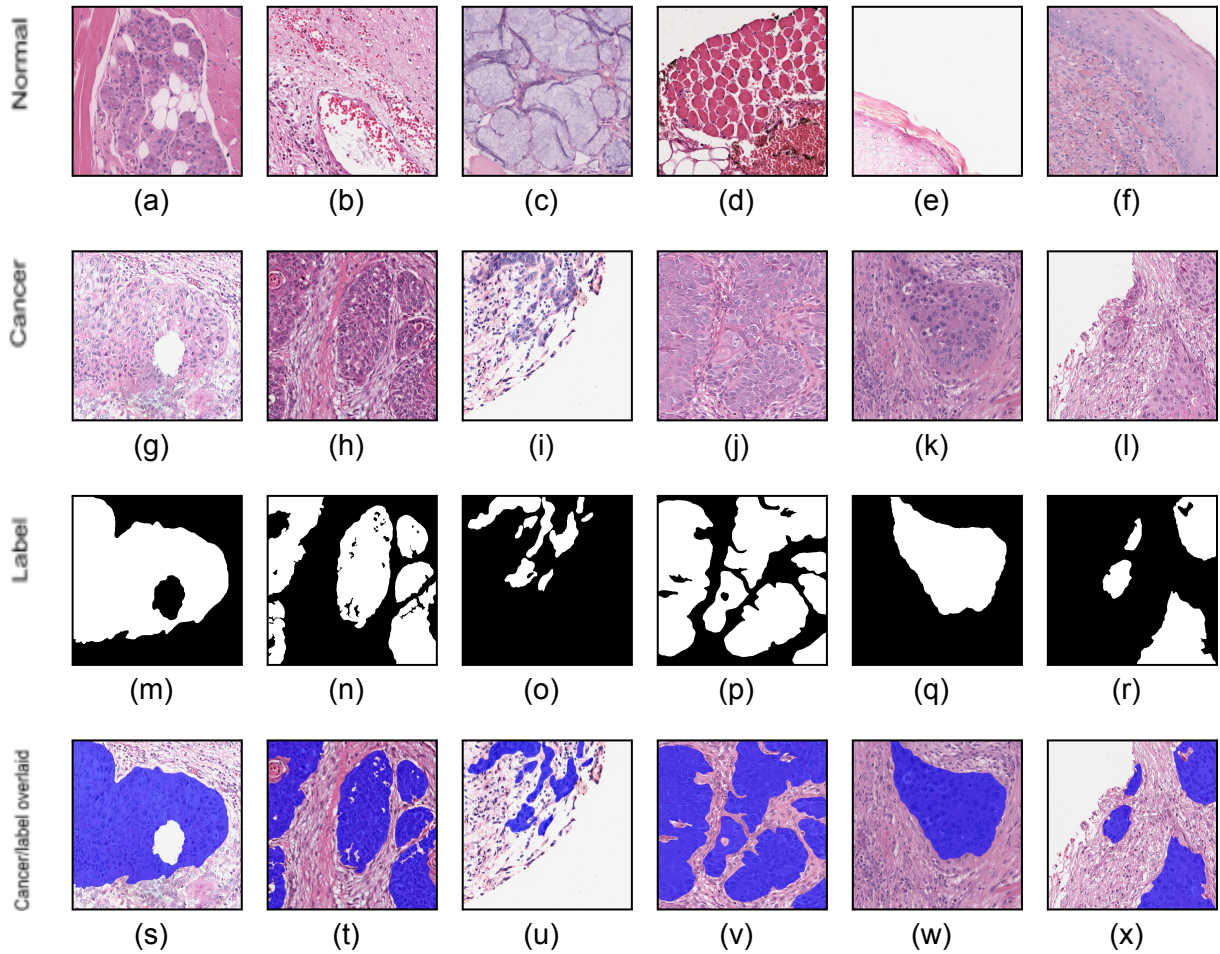

**Fig. 2.** Some images (640×640 pixels) from the OCDC dataset: 1st row (a-f) shows normal regions (serous salivary gland, connective tissue, mucous salivary gland, striated muscle, keratinized epithelial tissue, and oral mucosa); 2nd row (g-l) shows OSCC regions; 3rd row (m-r) shows the produced hand-annotated tumor regions (in white); 4th row (s-x) shows the cancer image with the identified tumor regions overlapped in blue.

**Ethics statements**
All OSCC cases were retrieved from the Department of Oral and Maxillofacial Pathology Archive at the Federal University of Uberlândia between 2006 and 2013 under approval by the Committee on Research and Ethics of the Institution (CAAE numbers: 15188713.9.0000.5152 and 58534122.7.0000.5152).

**Acknowledgments**
The authors gratefully acknowledge the financial support of National Council for Scientific and Technological Development CNPq (Grant 311404/2021-9) and the State of Minas Gerais Research Foundation - FAPEMIG (Grant APQ-00578-18).